\documentclass[a4paper, twocolumn, 11pt, accepted=2020-04-21]{quantumarticle}
\pdfoutput=1
\usepackage[utf8]{inputenc}
\usepackage[english]{babel}
\usepackage[T1]{fontenc}
\usepackage{amsmath,upgreek,amssymb,graphicx,subfigure,dsfont,color}
\usepackage{enumitem}
\setitemize{leftmargin=*}
\usepackage[sort&compress,numbers]{natbib}
\usepackage[colorlinks, linkcolor=quantumviolet, citecolor=quantumviolet, urlcolor=quantumviolet, breaklinks=true]{hyperref}

\usepackage{tikz}
\usepackage{lipsum}

\newcommand{\ket}[1]{|#1\rangle}
\newcommand{\bra}[1]{\langle#1|}
\newcommand{\braket}[1]{\langle#1\rangle}

\newcommand{\CSS}{\ket{\mathbf{\mathrm{x}}}}
\newcommand{\braCSS}{\bra{\mathbf{\mathrm{x}}}}
\newcommand{\charf}{X_A}
\newcommand{\evat}[1]{\big\rvert_{#1}}
\newcommand{\tr}[1]{\mathrm{tr}\left\lbrace #1 \right\rbrace}

\newcommand{\dephColl}{\gamma_C}
\newcommand{\dephIndiv}{\gamma_I}
\newcommand{\Lcol}{\mathcal{L}_C}
\newcommand{\Lind}{\mathcal{L}_I}

\newcommand{\colone}{blue}
\newcommand{\coltwo}{red}
\newcommand{\colthree}{black}

\begin{document}

\title{Ramsey interferometry with generalized one-axis twisting echoes}

\author{Marius Schulte}
\email{marius.schulte@itp.uni-hannover.de}
\orcid{0000-0003-0389-0988}
\affiliation{Institut f\"{u}r Theoretische Physik und Institut f\"{u}r Gravitationsphysik (Albert-Einstein-Institut), Leibniz Universit\"{a}t Hannover, Appelstra\ss e 2, 30167 Hannover, Germany}
%\author{Author2}
%\affiliation{Physikalisch-Technische Bundesanstalt, Bundesallee 100, 38116 Braunschweig, Germany}
\author{Victor J. Mart\'{i}nez-Lahuerta}
\affiliation{Institut f\"{u}r Theoretische Physik und Institut f\"{u}r Gravitationsphysik (Albert-Einstein-Institut), Leibniz Universit\"{a}t Hannover, Appelstra\ss e 2, 30167 Hannover, Germany}
\author{Maja S. Scharnagl}
\affiliation{Institut f\"{u}r Theoretische Physik und Institut f\"{u}r Gravitationsphysik (Albert-Einstein-Institut), Leibniz Universit\"{a}t Hannover, Appelstra\ss e 2, 30167 Hannover, Germany}
\author{Klemens Hammerer}
\orcid{0000-0002-7179-0666}
\affiliation{Institut f\"{u}r Theoretische Physik und Institut f\"{u}r Gravitationsphysik (Albert-Einstein-Institut), Leibniz Universit\"{a}t Hannover, Appelstra\ss e 2, 30167 Hannover, Germany}

\date{13.~March 2020}

\maketitle

\begin{abstract}
We consider a large class of Ramsey interferometry protocols which are enhanced by squeezing and un-squeezing operations before and after a phase signal is imprinted on the collective spin of $N$ particles. We report an analytical optimization for any given particle number and strengths of (un-)squeezing.
These results can be applied even when experimentally relevant decoherence processes during the squeezing and un-squeezing interactions are included. Noise between the two interactions is however not considered in this work. This provides a generalized characterization of squeezing echo protocols, recovering a number of known quantum metrological protocols as local sensitivity maxima, thereby proving their optimality. We discover a single new protocol. Its sensitivity enhancement relies on a double inversion of squeezing. In the general class of echo protocols, the newly found over-un-twisting protocol is singled out due to its Heisenberg scaling even at strong collective dephasing.
% [137/150 WORDS; NATURE PHYSICS ABSTRACTS LIMIT]
\end{abstract}

\section{Introduction}
Atomic sensors are currently among the most accurate measuring apparatuses in metrology, for e.g. precision spectroscopy, magnetometry or frequency metrology. Therefore, they may help to detect minute effects which can be indications of new physics~\cite{Ludlow_optical_2015, Safronova_Search_2018}. For high-precision atomic sensors, a fundamental limitation is given by quantum fluctuations of the measurement. However, entanglement between atoms allows these fluctuations to be reduced below what is possible with uncorrelated probes~\cite{Giovannetti_quantumMetrology_2006}. In the optical domain, squeezed states have been injected into laser interferometers to enhance precision~\cite{Barsotti_squeezed_2018, Tse_enhancedLIGO_2019, Acernese_increasingVIRGO_2019}. Likewise, significant successes followed in the generation and characterization of non-classical states in atomic physics~\cite{Pezze_quantumMetrology_2018}. However the biggest problems of these states are that strongly correlated systems often require demanding measurements and that increasing sensitivity typically comes at the cost of an increased susceptibility to imperfections. More detailed investigations, taking these effects into account, showed that the actual gain of such states can be significantly lower than in the ideal case~\cite{Huelga_improvement_1997, Escher_general_2011, DemkowiczDobrzaski_elusive_2012}. Hence the question of practical quantum metrology protocols arises. 

Recently, experiments in which simple but well-controlled interactions were used several times, in the form of an `echo', achieved excellent results in a variety of precision measurements~\cite{Hosten_quantum_2016, Leibfried_towards_2004, Linnemann_quantumEnhanced_2016, Burd_quantumAmplification_2019}. The one-axis-twisting (OAT) interaction for effective two-level systems~\cite{Kitagawa_squeezed_1993} allows a uniform description of many setups. It can be generated through cavity induced spin squeezing of cold atoms~\cite{SchleierSmith_states_2010, Pezze_quantumMetrology_2018}, via laser or microwave driven quantum gates for trapped ions~\cite{Blatt_entangled_2008} and from spin-changing collisions in spinor Bose-Einstein condensates~\cite{Pezze_quantumMetrology_2018}.

\begin{figure*}[tb]
	\centering
	\includegraphics[width=\linewidth]{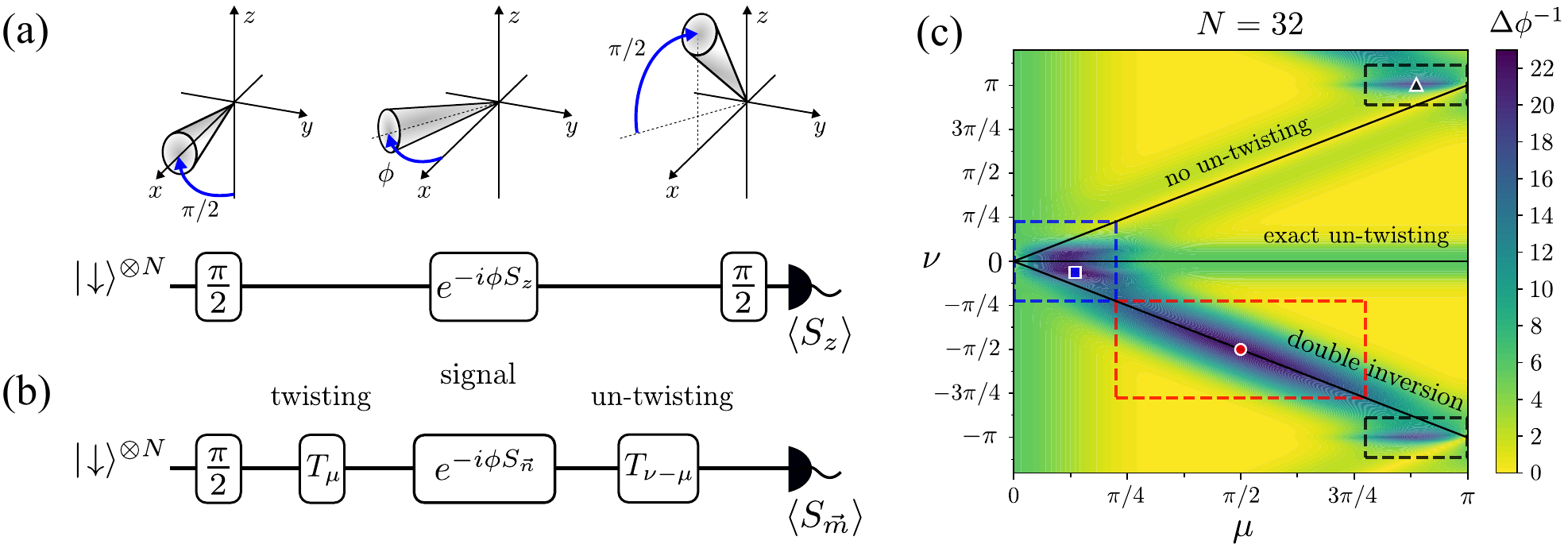}
	\caption{\textbf{Generalized Ramsey protocols:} (a) Conventional Ramsey measurement where first a phase $\phi$ is imprinted between two $\pi/2$-pulses by rotating the state in the equatorial plane around the $z$-axis. Finally, the signal in phase is determined from measuring the spin component $S_z$. The cones visualize the mean spin direction and quantum fluctuations of the state during the protocol.
	(b) Generalized Ramsey protocols with additional one-axis twisting $T_\mu$ and un-twisting $T_{\nu-\mu}$ dynamics as well as arbitrary directions $\vec{n}$, $\vec{m}$ for the signal and the measurement respectively. Optimizing over these additional degrees of freedom allows to extract the best sensitivity, characterized by the initial squeezing strength $\mu$ and the inversion $\nu$ only.
	(c) Example of an optimized sensitivity landscape for the inverse phase variance $\Delta \phi^{-1}$ (around $\phi=0$) with $N=32$. The boxes (dashed lines) highlight three distinct types of protocols we identified. At small $\mu$ the `squeezing protocols' (blue), and at large squeezing strength the `GHZ protocols' (black), which are known in the literature. In between, at an unusual double inversion of squeezing for $\nu = -\mu$, we find a new class of `over-un-twisting protocols' (red). Colored symbols denote the local maxima in each class.}
	\label{fig:landscape}
\end{figure*}

In our work we consider a large variational class of echo protocols based on OAT operations which, by construction, encompass a number of known protocols.  In order to study which protocols give useful enhancements, we also include collective and individual dephasing during the OAT interactions. Noise during the probe time, i.e. the application of the phase, is, however, beyond the scope of this work.
Related theoretical works include variational optimization algorithms~\cite{Kaubruegger_variational_2019} or inversion protocols using more general classes of spin-spin interactions which are referred to as interaction based readouts~\cite{Macri_loschmidt_2016, Haine_interactionBased_2018, Mirkhalaf_robustifying_2018, Anders_phase_2018, Huang_achieving_2018, Niezgoda_optimal_2019}. Here we obtain the maximal amplifications of signal-to-noise ratio (SNR) by analytically optimizing geometric parameters, that is, signal and measurement directions. This allows to provide a complete overview and classification of our echo protocols in terms of the squeezing and un-squeezing strengths (at any level of dephasing and arbitrary $N$). We identify one significant new scenario with a previously unused excess inversion. Such protocol types, which we refer to as `over-un-twisting' (OUT) protocols, are especially interesting as their sensitivity preserves the optimal Heisenberg scaling of entangled states, $\mathrm{SNR} \propto N$, independent of any collective dephasing during the OAT interactions. 
\section{Generalized Ramsey protocols}
For a long time now, measurements of small phase shifts were established as the framework for precision experiments in atomic and optical physics. In this regard, many areas of atomic physics adapted Ramsey interrogation~\cite{Ramsey_molecular_1950} due to the inherent reduction of systematic shifts.
The conventional Ramsey protocol is depicted in Fig.~\ref{fig:landscape}(a).
We consider $N$ two-level systems, described in terms of the collective spin operators $S_{x,y,z} = \frac{1}{2} \sum_{k=1}^N \sigma_{x,y,z}^{(k)}$ where $\sigma_{x,y,z}^{(k)}$ are the Pauli matrices for particle $k$.
Starting from the product state $\ket{\downarrow}^{\otimes N}$ with all particles in the ground state, a rotation 
\begin{equation}\label{eq:rotation}
    R_{\vec{n}} (\alpha) = e^{-i \alpha S_{\vec{n}}}
\end{equation}
around $\vec{n} = \vec{e}_y$ and with $\alpha=\pi/2$ generates the coherent spin state $\CSS = \left[\frac{\ket{\downarrow} + \ket{\uparrow}}{\sqrt{2}} \right]^{\otimes N}$, polarized in $x$-direction. (We use the notation $S_{\vec{n}} = \vec{n} \cdot \vec{S}$). In the following interaction period, a small phase signal $\phi$ is imprinted by rotating around the $z$-axis. A second $\pi/2$ rotation, with $\vec{n} = \vec{e}_x$, allows to infer $\phi$ from the measurement result of $S_z$.

%Idea: Extend classical Ramsey spectroscopy protocols with non-classical resources.
%Constrain this to allow only three actions: 
%First, rotations 
%\begin{equation}
%    R_{\vec{n}} (\alpha) = e^{-i \alpha S_{\vec{n}}}
%\end{equation}
%of the collective spin vector around an axis $\vec{n}$ with arbitrary angle $\alpha$. Here $S_{\vec{n}} = \vec{n} \cdot \vec{S}$.
%Second, the non-linear one axis twisting (OAT) interaction
%\begin{equation}
%    T_{\mu} = e^{-i \frac{\mu}{2} S_z^2}
%\end{equation}
%generated physically by applying the quadratic Hamiltonian $H = \chi S_z^2$ for a time $t$, such that $\chi t = \mu/2$.
% Finally we consider measurements of the mean collective spin $S_{\vec{m}}$ in an arbitrary direction $\vec{m}$.

Motivated by the conventional Ramsey scheme, we now consider the generalized echo protocols shown schematically in Fig~\ref{fig:landscape}(b).
Starting again from the coherent spin state $\CSS$ after the first $\pi/2$ pulse, a squeezed spin state is generated via the non-linear OAT interaction $T_{\mu} = e^{-i \frac{\mu}{2} S_z^2}$, with strength $\mu$. The necessary Hamiltonian $\propto S_z^2$ can be engineered in a variety of metrologically relevant systems~\cite{Pezze_quantumMetrology_2018, Blatt_entangled_2008}. For small squeezing strength, $\mu < 4/\sqrt{N}$, the generated entanglement is reflected by reduced fluctuations of the mean spin~\cite{Kitagawa_squeezed_1993, Pezze_quantumMetrology_2018}. With increasing $\mu$ greater levels of multi-particle entanglement are generated. 
At $\mu = \pi$ rotated versions of the $N$-particle correlated Greenberger-Horne-Zeilinger (GHZ) state $\left[\ket{\downarrow}^{\otimes N} + \ket{\uparrow}^{\otimes N}\right]/\sqrt{2}$, aligned along the $x$ axis if $N$ is even or along the $y$ axis if $N$ is odd, are created and the dynamics reverses afterwards~\cite{Pezze_quantumMetrology_2018}.
For the following signal imprint we now allow the more general case of a rotation of the collective spin around an arbitrary axis $\vec{n}$. Note that a physical rotation around the $z$-axis may be converted to a rotation around $\vec{n}$ by appropriate single qubit rotations $R$ (in the form of Eq.~\eqref{eq:rotation}) before and after the phase imprint such that $R_{\vec{n}}(\phi)=e^{-i \phi S_{\vec{n}}} = R^{-1} e^{-i \phi S_{z}} R$.
Prior to the measurement we allow for another OAT interaction with strength $\nu - \mu$. With this definition, $\nu$ describes the deviation from an exact inversion of the initial OAT. This choice is based on an appearing symmetry around exact echo protocols at $\nu = 0$, c.f. Fig.~\ref{fig:landscape}(c). We assume inversion of OAT is possible by reverting the sign of the interaction strength, as already demonstrated for cold atoms~\cite{Hosten_quantum_2016}, spinor Bose-Einstein condensates~\cite{Linnemann_quantumEnhanced_2016} and trapped ions~\cite{Leibfried_towards_2004, Gaerttner_measuringOTOC_2017}.
At the end of our protocols the collective spin $S_{\vec{m}}$ in an arbitrary direction $\vec{m}$ is measured. Again, this can be implemented with a measurement of $S_z$ by preceding an appropriate rotation of the collective spin.
Overall, the generalized echo protocols have a measurement signal
\begin{equation}
    \langle S_{\vec{m}} \rangle (\phi) = \braCSS T_{\mu}^{\dagger}\, R^{\dagger}_{\vec{n}} (\phi)\, T_{\nu-\mu}^{\dagger} \, S_{\vec{m}} \, T_{\nu-\mu} \, R_{\vec{n}} (\phi)\, T_{\mu} \CSS
\end{equation}
characterized by parameters $\mu, \nu$ for squeezing and un-squeezing and directions $\vec{n}, \vec{m}$ for signal and measurement.

%%%%
% Two main ideas for improving the SNR by spin squeezing developed: On the one hand, reducing the quantum projection noise at constant signal strength with an initially squeezed state and on the other hand the amplification of the signal at constant noise by an additional inversion of the squeezing before the final spin measurement.
% The noise minimization of the first method finds its limitation in the loss of contrast for strongly squeezed states and is therefore constrained to moderate squeezing only.
% However efficient applications of strongly squeezed states, e.g. in atomic clocks, exist but require more complex adaptive strategies involving weak measurements of the collective spin.
% On the other hand, the second protocol has the disadvantage that the strong signal amplification leads to a reduced measurement range (bandwidth).
% This raises the question of whether there are intermediate methods, with partial squeezing inversion, which avoid the above disadvantages.
%%%%

% At this point, 
We highlight that the generalized echo protocols include some common squeezing protocols as limiting cases: For $\nu = \mu$, with no un-twisting, we find standard Ramsey interrogation with a spin squeezed initial state~\cite{Andre_stability_2004}. In this case the SNR is enhanced by reducing projection noise at a constant signal. More recently, protocols with exact inversion, i.e. $\nu =0$, were suggested for application in quantum metrology~\cite{Davis_approaching_2016, Froewis_detecting_2016, Nolan_optimal_2017}. There, amplification of the signal at constant measurement noise occurs. 
% Beyond these known approaches, new protocols are also discovered from our theory as we will show.
\section{Geometric optimization}
In the following we quantify metrological gain by the inferred phase deviation
\begin{equation}\label{eq:FoM}
    \Delta \phi (\mu, \nu, \vec{n}, \vec{m}) = \Delta S_{\vec{m}}\vert_{\phi = 0}  \bigg / \left\vert \frac{\partial \langle S_{\vec{m}}\rangle}{\partial \phi}\big\rvert_{\phi =0} \right\vert
\end{equation} 
around the working point $\phi = 0$. This is a measure of sensitivity which describes enhancements in atomic sensors limited by quantum projection noise.
We will now show that the essential optimization with respect to the signal and measurement directions can be solved analytically. The method described in the following corresponds exactly to the method first described by Gessner et al. in~\cite{Gessner_metrological_2019}.
While in that case the authors were able to systematically optimize measurement operators, we use the method in our work to find the optimal geometric factors for generalized echo protocols. It is interesting to note that only by this means the metrological gain of strongly entangled states, such as non-Gaussian states, could be shown in both works.

First, we re-express the two contributions, signal and noise, separately. For the signal strength we find
\begin{equation}\label{eq:slope}
    \left\vert \frac{\partial \langle S_{\vec{m}}\rangle}{\partial \phi}\big\rvert_{\phi =0} \right\vert = \vec{n}^T M \vec{m}
\end{equation}
with $M_{kl} = i \langle [S_k(\mu), S_l(\nu) ] \rangle_{\vert_{\phi = 0}}$ where we denoted transformed spin operators by $S_{\vec{n}}(\mu) = T_{\mu}^{\dagger} S_{\vec{n}} T_{\mu}$.
Likewise, the measurement variance can be expressed as
\begin{equation}
    \Delta S^2_{\vec{m}}\,_{\vert_{\phi = 0}} = \langle S^2_{\vec{m}}(\nu) - \langle S_{\vec{m}}(\nu) \rangle^2 \rangle_{\vert_{\phi = 0}} = \vec{m}^T Q \vec{m}
\end{equation}
with $Q_{kl} = \langle S_{k}(\nu) S_{l}(\nu) - \langle S_{k}(\nu) \rangle \langle S_{l}(\nu) \rangle \rangle_{\vert_{\phi = 0}}$.
The matrices $M$ and $Q$ can be obtained analytically (see appendix~\ref{app:CharF}).

For sensitivity optimization we aim to maximize the inverse phase deviation, i.e. SNR, which is expressed as
\begin{equation}
    \Delta \phi^{-1} = \frac{\vec{n}^T M \vec{m}}{\sqrt{\vec{m}^T Q \vec{m}}}\,.
\end{equation}
Because $Q$ is a positive semi-definite spin co-variance matrix, we can define the vector $\vec{v} = Q^{1/2} \vec{m}$ with $\vert\vert \vec{v}\vert \vert = \sqrt{\vec{m}^T Q \vec{m}}$ and correspondingly $\vec{m} = Q^{-1/2} \vec{v}$. Note that $Q$ is singular for $\nu = 0$ only. In this case, the optimization of rotation and measurement directions can be restricted to the plane perpendicular to the initial spin polarization.
With the unit vector $\vec{u} = \vec{v}/\vert\vert \vec{v}\vert\vert $ the sensitivity is
\begin{equation}
    \Delta \phi^{-1} = \vec{n}^T\, M\, Q^{-1/2}\, \vec{u}
\end{equation}
This can be optimized over all signal directions $\vec{n}$ and measurement directions $\vec{m}$ by a singular value decomposition for $ M\, Q^{-1/2}$.
After this step the sensitivity depends exclusively on the initial squeezing strength $\mu$ and excess inversion $\nu$. The ideal directions for signal and measurement, at each point $(\mu, \nu)$, can be inferred from two orthogonal matrices which are determined by the singular value decomposition.
An example landscape of the optimal SNR, i.e. maximal singular value,
\begin{equation}
    \Delta \phi^{-1} (\mu, \nu) = \max_{\vec{n}, \vec{m}} \Delta \phi^{-1} (\mu, \nu, \vec{n}, \vec{m})
\end{equation}
is shown in Fig.~\ref{fig:landscape}(c).
We point out that the relatively small particle number, $N=32$, there is motivated to easily highlight important features of the landscape. With the analytic expressions, computational time is independent of $N$ and we show landscapes for larger particle numbers in the appendix, Fig.~\ref{fig:all_landscapes}.

\begin{figure*}[tb]
	\centering
	\includegraphics[width=\linewidth]{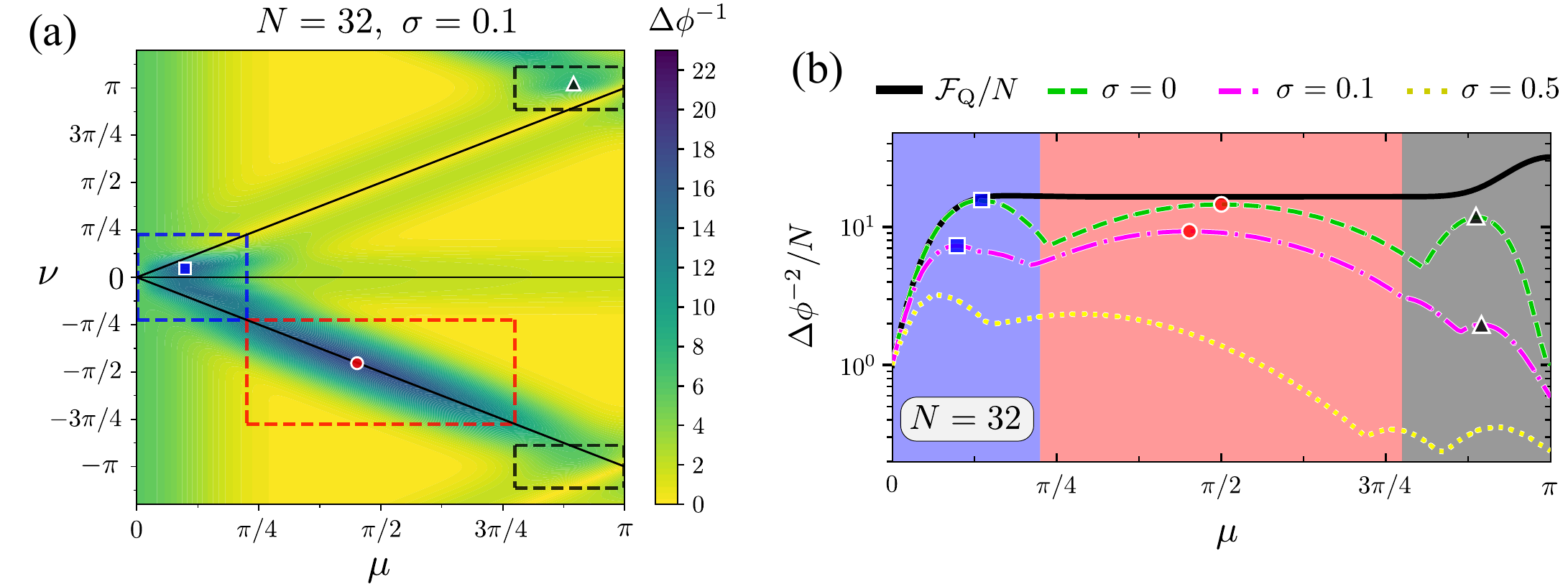}
	\caption{\textbf{Sensitivity with dephasing:} (a) Sensitivity $\Delta\phi^{-1}$ for $N=32$ with $\sigma = 0.1$ collective dephasing noise during the state preparation and inversion. The overall structure of generalized echo protocols remains intact. As compared to Fig.~\ref{fig:landscape}(c), changes in the positions of optimal protocols (symbols) as well as an overall reduction in sensitivity can be observed. (b) SNR with optimized inversion, $\max_{\nu} \Delta\phi^{-2}(\mu, \nu)/N$, for different levels of collective dephasing noise $\sigma=0,\,0.1,\,0.5$. Colored areas mark the squeezing, OUT and GHZ protocol types. The black line is the scaled quantum Fisher information $\mathcal{F}_Q/N$ of the ideal initial squeezed state, i.e. $\sigma=0$. We find three peaks corresponding to the optimal squeezing, over-un-twisting and GHZ protocols. With increasing particle number, only the over-un-twisting protocols are robust to small collective dephasing with a sensitivity close to the fundamental quantum Fisher information limit. Symbols on the green dashed line represent the optimal protocols of Fig.~\ref{fig:landscape}(c) while symbols on the magenta dash-dotted line correspond to part (a).}
\label{fig:noise}
\end{figure*}

In view of Fig.~\ref{fig:landscape}(c) mainly three separate regions exist in which amplified SNR is achieved: \textit{(i)} The first region (\colone~box) includes protocols with small squeezing strengths $\vert \mu\vert, \vert\nu\vert \lesssim 4/\sqrt{N}$. We refer to such cases as `squeezing protocols' because this is the only region that contains initial states exhibiting reduced spin fluctuations while still maintaining significant polarization. Note however that this usual intuition is no longer exclusive for all states in this class. Already at $\mu \gtrsim 2/\sqrt{N}$ the states generated by OAT enter the regime of so called oversqueezed states, which wrap around the Bloch sphere. From this point on spin squeezing is lost, i.e. $\xi >1 $, according to the Wineland squeezing parameter $\xi$~\cite{Wineland_spinSqueezing_1992, Wineland_squeezed_1994}. This is connected to our figure of merit without any echo ($\nu =\mu$) by $\xi = N \Delta \phi^2(\mu, \mu)$. While the original classification based on the squeezing parameter $\xi$ fails with the application of echoes, we find that the characteristic scaling $\sim 1/\sqrt{N}$ with particle number remains valid for the generalized protocols. Note that the failure of this argument is also visible in the landscapes we show. Compare e.g. the position of the maximum along the $\nu = \mu$ line with the local maximum over all squeezing protocols (blue square) in Fig.~\ref{fig:landscape}(c). The additional factor of two in the condition $\vert \mu\vert, \vert\nu\vert \lesssim 4/\sqrt{N}$ was introduce here to place the cutoff close to the minimum that lies between the local maximum at small squeezing strength and the broad maximum around $\mu = \pi/2$, cf. Fig.2(b). 

As a special case of `squeezing protocols' we recover the classic Ramsey protocols with squeezed initial states and no un-squeezing ($\nu = \mu$) along with their known optimal signal and measurement directions~\cite{Kitagawa_squeezed_1993}. 
We also find optimized exact echo protocols, on the horizontal line at $\nu = 0$, for initial squeezing strengths $\mu \sim 2/\sqrt{N}$. The optimal directions converge to $\vec{n} = \vec{e}_y$ and $\vec{m} = \vec{e}_y$ for $N \gg 1$. This proves optimality of the choices of signals and measurements made by Davis et al.~\cite{Davis_approaching_2016}. Interestingly, we find that within the class of squeezing protocols the local maximum in sensitivity is reached at values of $\nu$ which do not correspond to either of these known protocols. \textit{(ii)} The second region (\colthree~box) exhibiting enhanced sensitivity exists at large squeezing strength, $\pi - 4/\sqrt{N} \leq \mu \leq \pi$, which contains highly entangled states close to the GHZ state, so that we refer to these as `GHZ protocols'. 
Note that the enhancement of GHZ states is ideally obtained with parity measurements. Recently, approaches using the measurement of spin projections and an additional squeezing interaction were discussed by Leibfried et al.~\cite{Leibfried_towards_2004, Leibfried_creation_2005}. \textit{(iii)} Finally, we find exactly one more region (\coltwo~box, with $ 4/\sqrt{N} < \mu < \pi - 4/\sqrt{N}$) corresponding to a new type of protocols that are characterized by a double inversion of the OAT, at $\nu=-\mu$. We refer to these as `over-un-twisting' (OUT) protocols. The enhancing mechanism there is conceptually different from the squeezing protocols and is discussed further below.
The initial states in this class are regarded as oversqueezed or non-Gaussian states~\cite{Strobel_fisherInformation_2014}. So far the entanglement was first captured by the quantum Fisher information~\cite{Pezze_entanglement_2009} or only in terms of nonlinear squeezing parameters~\cite{Gessner_metrological_2019}.

% The prospect to infer the metrological gain of GHZ states via repeated entangling interactions and measurements of spin projections rather than a direct parity measurement was already discussed by Leibfried et al.
% First observations of the increased sensitivity using up to six trapped ions were shown in the same references.
% However, the GHZ protocols have the drawback that their enhancement is limited much stronger by noise during the OAT state preparation and inversion than the squeezing or OUT protocols.

% In the first case the amplification of small phases can typically be explained by either the increased rotation of an almost coherent spin state or the reduction of measurement noise as suggested above. 

\section{Dephasing noise}
% \noindent \textit{-- Dephasing during OAT interactions}
% We now extend the above discussion of sensitivity by including dephasing noise during the initial OAT interaction and inversion.
To see which protocols actually correspond to a robust enhancement, we now add dephasing during the OAT.
In the presence of collective dephasing, at a rate $\dephColl>0$, the dynamics of the system will be governed by the master equation
\begin{equation}\label{eq:masterEq}
    \dot{\rho} = -i [H, \rho] + \dephColl \Lcol [\rho]
\end{equation}
with $H = \chi S_z^2$ and $\Lcol [\rho] = S_z \rho S_z - \frac{1}{2} S_z^2 \rho - \frac{1}{2} \rho S_z^2$.
%In both cases the noise commutes with the unitary evolution, the formal solution to the master equation is
%\begin{equation}\label{eq:solMasterEq}
%    \rho (t) = T_{\mu} e^{\sigma \frac{\vert \mu \vert}{2} \mathcal{L}_k}[\rho_0] T_{\mu}^{\dagger} = e^{\sigma \frac{\vert \mu \vert}{2} \mathcal{L}_k}[T_{\mu} \rho_0 T_{\mu}^{\dagger}]
%\end{equation}
%with $\mu = 2 \chi t$ as before and $\sigma = \vert \dephColl \vert / \vert \chi \vert$ quantifies the noise in terms of the OAT interaction strength.
The noise strength is quantified by the dimensionless parameter $\sigma = \vert \dephColl \vert / \vert \chi \vert$. 
In cavity induced squeezing of atoms such dephasing occurs as fluctuations in the phase of the collective spin due to photon shot noise~\cite{Davis_approaching_2016}. For quantum gates with trapped ions, dephasing occurs through random variations of the transition frequency from stray fields or from frequency noise of the driving fields. When uniform over the extend of the ion string, both result in collective dephasing~\cite{Kielpinski_decoherenceFree_2001, Roos_designer_2006}.
With spinor Bose-Einstein condensates, collective dephasing may again arise from magnetic field fluctuations~\cite{Pezze_quantumMetrology_2018}.

% In appendix~\ref{app:noise} we show how the expectation values contained in $M$ and $Q$ can be related to their noiseless values and we state the explicit matrices.

%So including dephasing noise results in 
%\begin{equation}
%    \tilde{M} = \begin{pmatrix}
%    \frac{1}{2} e^{-\sigma \frac{\vert \nu - \mu \vert}{4}} [n_1 + e^{-\sigma \vert \mu \vert} n_2] & 0 & 0 \\
%    0 & \frac{1}{2} e^{-\sigma \frac{\vert \nu - \mu \vert}{4}} [n_1 - e^{-\sigma \vert \mu \vert} n_2] & e^{-\sigma \vert \mu \vert/4} n_3 \\
%    0 & e^{-\sigma \frac{\vert \nu - \mu \vert + \vert \mu \vert}{4}} n_4 & 0
%    \end{pmatrix}
%\end{equation}
%and
%\begin{equation}
%    \tilde{Q} = \begin{pmatrix}
%    \frac{1}{2} [q_1 + e^{-\sigma (\vert \nu - \mu \vert + \vert \mu \vert)} q_2] - e^{-\sigma \frac{\vert \nu - \mu \vert}{2}} q_0^2 & 0 & 0 \\
%    0 & \frac{1}{2} [q_1 - e^{-\sigma (\vert \nu - \mu \vert + \vert \mu \vert)} q_2] & e^{-\sigma \frac{\vert \nu - \mu \vert}{4}} q_3 \\
%    0 & e^{-\sigma \frac{\vert \nu - \mu \vert}{4}} q_3 & q_4
%    \end{pmatrix} .
%\end{equation}

\begin{figure}[tbp]
	\centering
	\includegraphics[width=\linewidth]{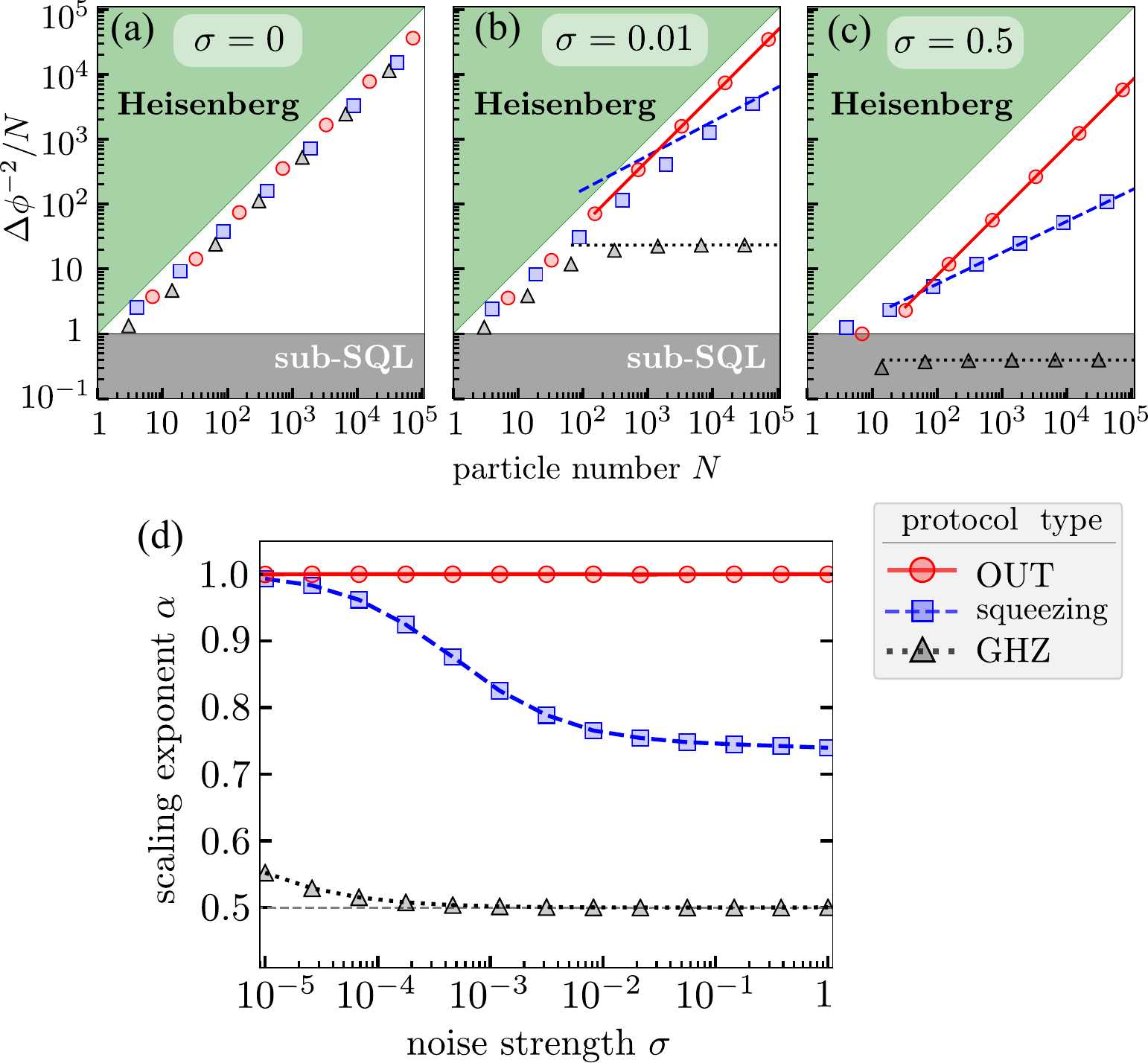}
	\caption{\textbf{Scaling of sensitivity with particle number:} (a)-(c) Scaling of the sensitivity under collective dephasing with $\sigma= 0.01, 0.1, 0.5$ for squeezing (blue), OUT (red) and GHZ (black) protocols. The green region shows sensitivity beyond the Heisenberg limit, $\Delta \phi^{-2} > N^2$, and the grey region sensitivity below the standard quantum limit, $\Delta \phi^{-2} < N$. Lines are fits of the local optima to asymptotic scalings $\Delta \phi^{-1} = c\,N^{\alpha}$ at large $N$. Note that a flat line thus still means improvement with $N$, however only at the classical scaling $\propto N$. (d) Scaling exponent $\alpha$ as a function of noise strength for the three protocol types identified above.}
	\label{fig:scaling}
\end{figure}

The geometrical optimization with respect to $\vec{n}$ and $\vec{m}$ can again be performed analytically, see appendix~\ref{app:noise}. Figure~\ref{fig:noise}(a) shows the sensitivity including collective dephasing with $\sigma=0.1$.
Compared to the ideal case, shown in Fig.~\ref{fig:landscape}(c), we see that any enhancement of the GHZ protocols is strongly suppressed by the noise. Furthermore, within the squeezing protocols a bias towards $\nu \approx \mu$ is developing as protocols with reduced additional inversion suffer less from dephasing.
Surprisingly, the large maximum of OUT protocols remains only weakly affected by preparation noise and still offers large enhancement. To emphasize this, Fig.~\ref{fig:noise}(b) displays the measurement optimized sensitivity $\max_{\nu} \left(\Delta\phi^{-2}(\mu, \nu)/N \right)$ as a function of the initial squeezing at various levels of dephasing.
% We find that this distinguishes well between the emerging locally optimal spots even up to larger values.
% This shows that the sensitivity of OUT protocols is only weakly reduced even up to moderate dephasing strengths around $\sigma \sim 0.1$, while the reduction in sensitivity of squeezing and GHZ protocols is much stronger.
Performing the optimization over all measurements within our protocols also allows to compare the obtained SNR to fundamental limits of quantum metrology.
% , characterized by $\nu \in [-\pi, \pi]$,
% as functions of $\mu$
% Optimizing in this way over all parameters except the initial squeezing strength allows to compare our obtained SNR to more fundamental limits.
Most notably, the quantum Cram\'{e}r-Rao bound implies $ \max_{\nu} \Delta\phi^{-2}(\mu, \nu) \leq \mathcal{F}_Q$.
The upper limit $\mathcal{F}_Q$ is the quantum Fisher information, which quantifies the maximum information about the phase $\phi$ that can possibly be inferred from the initial squeezed state under any rotation $S_{\vec{n}}$.
The quantum Fisher information thereby includes an optimization over all measurements, including weak measurements, individual operations on each particle, parity and others, which go beyond what is possible with the resources considered here.
Based on~\cite{Pezze_entanglement_2009} we show the quantum Fisher information limit (black line) in Fig.~\ref{fig:noise}(b) in comparison to the SNR.
% Note that there the quantum Fisher information is additionally maximized over all rotations $R_{\vec{n}}(\phi)$ imprinting the phase signal (see appendix~\ref{app:QFI} for details).
As a function of the squeezing strength, $\mathcal{F}_{\mathrm{Q}}$ increases from the standard quantum limit $\mathcal{F}_{\mathrm{Q}} = N$ of uncorrelated particles at $\mu = 0$ up to the Heisenberg limit $\mathcal{F}_{\mathrm{Q}} = N^2$ at $\mu = \pi$.
Even though the quantum Fisher information constitutes a true extension over the capabilities of the protocols considered here, we nevertheless find that the OUT protocols actually reach the quantum Fisher information bound with increasing $N$. This feature persists for small dephasing as well. The only other case where this holds true is for $\mu \ll 1$. However, at extreme levels of noise also the OUT protocols fall short compared to the quantum Fisher information limit.
% , thus showcasing the insensitivity of OUT protocols to small imperfections.
% which is not surprising as these protocols also involve significant interactions before the measurement

Due to the exact optimization introduced above we are also able to efficiently examine the influenced of dephasing on the particle number scaling of the sensitivity.
Figure~\ref{fig:scaling}(a)-(c) show the scalings for $\sigma = 0, 0.01, 0.5$. Symbols mark the best sensitivity within each protocol type while lines show numerical fits to an asymptotic scaling $ \Delta \phi^{-1} = c N^{\alpha}$ with fitting parameters $c, \alpha$. The green region reflects sensitivity beyond the Heisenberg limit $ \Delta \phi^{-2} > N^2$ and the grey region sensitivity below the standard quantum limit $ \Delta \phi^{-2} < N$. 
% Note that thus also a flat line in Fig.~\ref{fig:scaling}(a)-(c) corresponds to an increased sensitivity with $N$, but only at the classical scaling. 
We find that, remarkably, the OUT protocols always exhibit Heisenberg scaling, $\Delta \phi^{-1} \propto N$, independent of the dephasing. On the other hand the GHZ protocols quickly drop to the classic scaling, showcasing their increased susceptibility in this regard. The dependence of the exponent $\alpha$ on the noise strength is shown in Fig.~\ref{fig:scaling}(d), highlighting the characteristic differences regarding the influence of collective dephasing.
Although the squeezing protocols have a reduced scaling exponent compared to OUT protocols, they may still be the overall best protocol when limited to small ensembles and larger dephasing, cf. Fig.~\ref{fig:scaling}(c).
Our findings also show that initially the scaling may deviate significantly from the asymptotic case, even up to ensembles of considerable size.

% This is due to the $\sigma$ dependence of the pre-factor $c$.
% Besides the exponent we find that especially for small ensembles $N \lesssim 100$ the pre-factor plays a decisive role, whereby the protocols with \textit{low squeezing} often achieve better sensitivity in this regime.

\begin{figure}[tb]
	\centering
	\includegraphics[width=\linewidth]{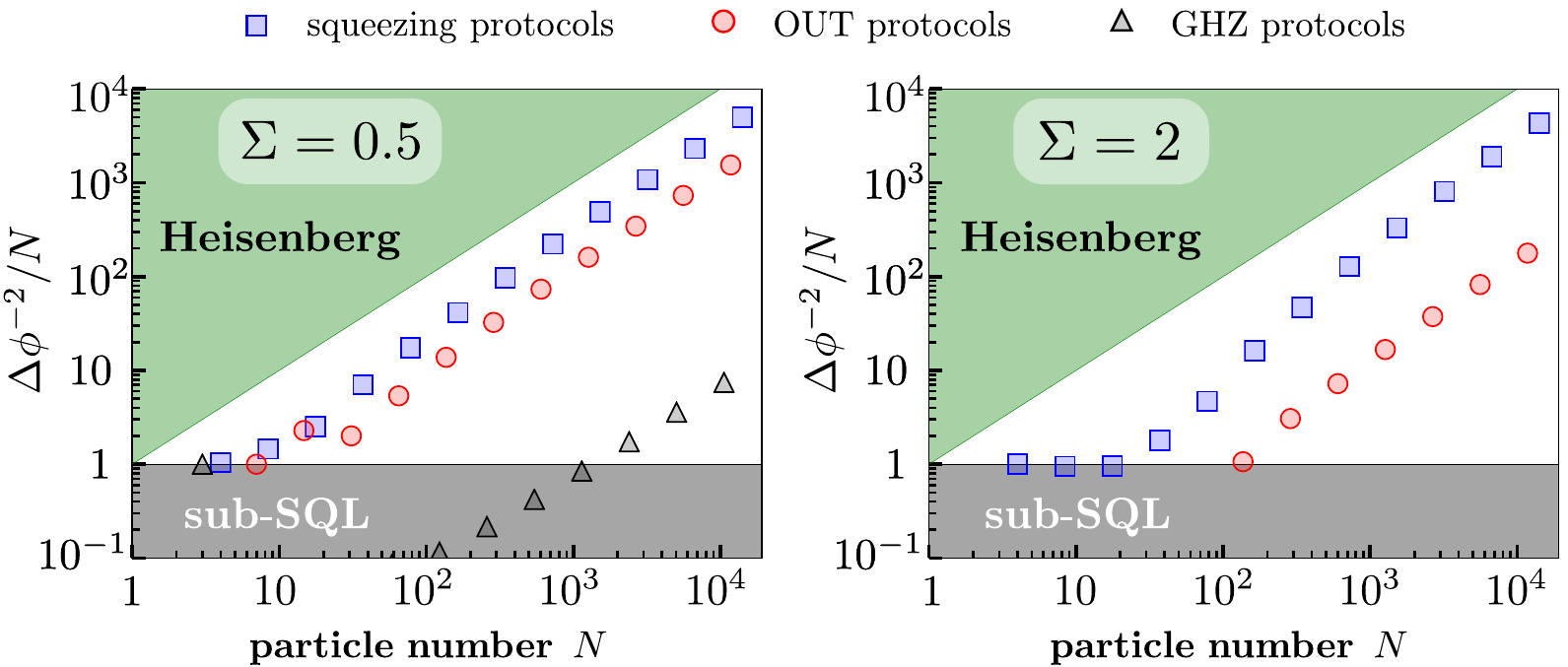}
	\caption{\textbf{Effects of individual dephasing:} Sensitivity versus particle number for individual dephasing during the OAT interactions with strength $\Sigma = 0.5$ and extreme dephasing at $\Sigma = 2$.}
	\label{fig:Indiv_scaling}
\end{figure}

In addition to collective dephasing we also study individual dephasing during the OAT interactions.
Compared to Eq.~\eqref{eq:masterEq}, the master equation is $\dot{\rho} = -i [H, \rho] + \dephIndiv \Lind [\rho]$, where $\Lind[\rho] = \sum_{k=1}^N \sigma_z^{k} \rho \sigma_z^{k} - \rho$. This describes individual but symmetric dephasing of each atom at a rate $\dephIndiv>0$ and we likewise define $\Sigma = \vert \dephIndiv \vert / \vert \chi \vert$. It turns out that individual dephasing results in a less stringent restriction on sensitivity than collective dephasing. The sensitivity for all protocol types scales asymptotically linearly in $N$, independent of the noise strength, as shown in Fig.~\ref{fig:Indiv_scaling}.
% at the same noise strength

\begin{figure*}[tb]
	\centering
    \includegraphics[width=\linewidth]{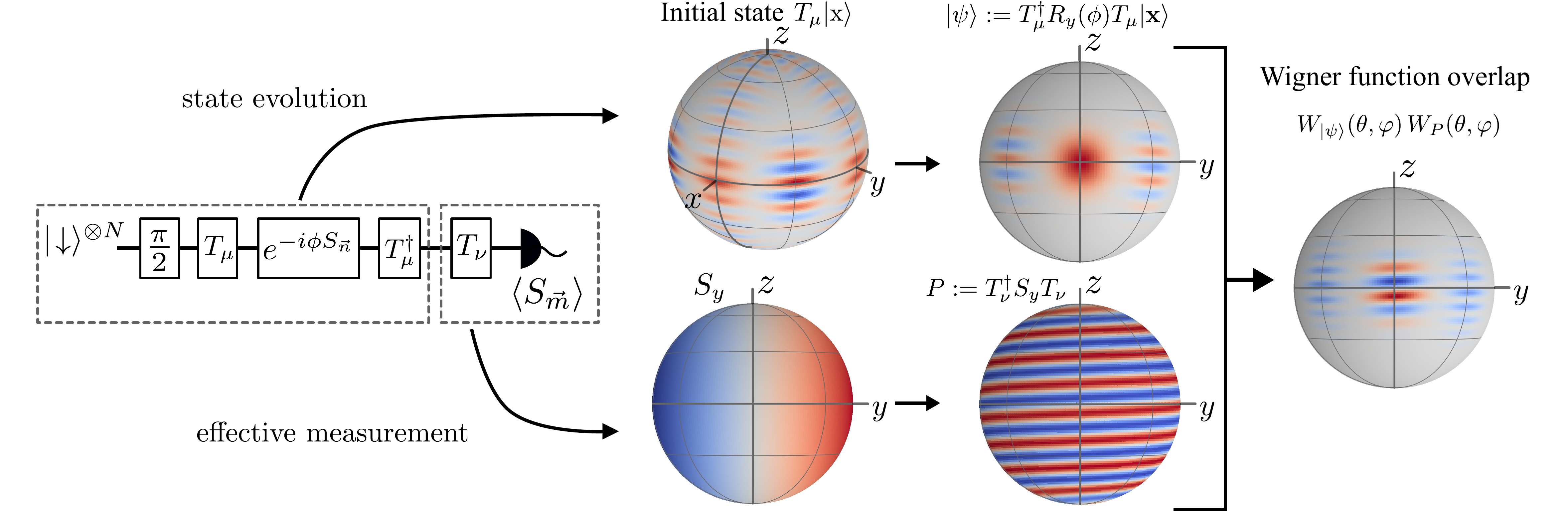}
	\caption{\textbf{Illustration of over-un-twisting protocols:} Wigner function representation for the optimal over-un-twisting protocol at $\mu = \frac{\pi}{2}$, $\nu = -\frac{\pi}{2}$ and with $N=32$. Half the OAT inversion is assigned to the state evolution, the other half to the measurement operator. State evolution (top): Starting from the superposition of four coherent states along the equator, a small rotation around $y$ (here $\phi = -0.02$) followed by exact OAT inversion, leads to the state $\ket{\psi}$ with large polarization in $x$-direction and residual interference patterns due to the disturbance of the rotation. Effective measurement (bottom): OAT of the optimal measurement direction $\approx S_y$ creates dense sequences of positive and negative values in its Wigner function. These match exactly the pattern on $\ket{\psi}$, so that in the overlap of the two Wigner functions the interference patterns are either all positive or all negative, depending on the sign of $\phi$. Integrating the overlap over the sphere results in the measurement signal $\braket{S_y}$. Larger values of $N$ have narrower spacing of the interference patterns with increased maximal and minimal values of the Wigner function, thus giving an enhanced signal.}
	\label{fig:double_inversion}
\end{figure*}

\section{Over-un-twisting enhancement}
In this last section we give an explanation of the mechanism underlying the OUT protocols.
To visualize the sensitivity enhancement we split the protocols into a state evolution and an effective measurement part.
As shown in the schematic of Fig.~\ref{fig:double_inversion} we group one half of the double inversion to the state evolution and the other half to the measurement.
Now for the optimal OUT protocol, initially a superposition of four coherent spin states along the equator of the collective Bloch sphere is generated by $T_{\mu}$ with $\mu=\frac{\pi}{2}$~\cite{Agarwal_atomicSchroedingerCat_1997}, as shown by the Wigner function~\cite{Dowling_wigner_1994} in the top row of Fig.~\ref{fig:double_inversion}.

A small rotation around the $y$-axis perturbs the following exact inversion of OAT in such a way that interference patterns remain on the sides of a large polarization contribution in $x$-direction.
%~\footnote{There are two almost equal singular values, the larger one with the $y$-axis for rotation and measurement and the slightly smaller one with the $x$-axis, which becomes clear from the symmetry of the state}. 
The absolute values of these patterns increase for larger rotation angles $\phi$ or with increasing $N$ when keeping $\phi \ll 1$ fixed. The signs of the interference patterns change only when rotating in the opposite direction, i.e. $R_y(-\phi)$ instead of $R_y(\phi)$.
The second part of the OUT protocols can be viewed as a transformation of the applied measurement. The bottom row in Fig.~\ref{fig:double_inversion} shows that the twisting dynamics on $S_y$ leads to a Wigner function for the operator $P:= T_{\pi/2}^{\dagger} S_y T_{\pi/2}$ with rapid sign changes, wrapping around the Bloch sphere. Note that the density of patterns increases in the same way with $N$ for both the state $\ket{\psi}$ and the measurement operator $P$. So importantly the interference patterns of the two Wigner functions match exactly. 
It is interesting to note that the Wigner functions show similarities to a Moir\'{e} pattern as well as Gottesman-Kitaev-Preskill states~\cite{Gottesman_encoding_2001, Duivenvoorden_single-mode_2017}. Due to the matching Wigner functions, in the product all oscillations contribute with either a positive or a negative sign, depending on the sign of the rotation. The mean $\braket{S_y} = \int_0^{\pi} \mathrm{d}\theta \int_0^{2 \pi} \mathrm{d}\varphi \sin (\theta) W_{\ket{\psi}}(\theta ,\varphi) W_{P}(\theta ,\varphi) $ then corresponds to the integral over the sphere for the product of the Wigner functions~\cite{Agarwal_relation_1981, Dowling_wigner_1994}.
This shows $N$ times faster oscillations, compared to uncorrelated atoms, for even $N$. For odd $N$ the signal has a sharp initial incline before vanishing after a few oscillations (see Fig.~\ref{fig:wigner_functions} in the appendix).
Although this distinction will be relevant for larger phases $\phi$, both the figure of merit in Eq.~\eqref{eq:FoM} and the optimal signal and measurement directions remain a continuous function of $N$. Thus, for the OUT protocols no additional information about the particle number is necessary. This is a consequence of the fact that we study and optimize the local sensitivity around $\phi = 0$.
Beyond the local sensitivity the mean information obtained per estimate can be calculated following~\cite{Hall_nonlinear_2012}. We find that the information is similar to what is achieved with coherent spin states, in stark contrast to GHZ states which return only a single bit of information. The fact that we find an increased information gain, as well as a large local sensitivity is a compelling property of this novel class of protocols.

% Additionally, the OUT protocols also show similarities to sensing schemes with Gottesman-Kitaev-Preskill states and modular variable measurements as proposed for continuous variable systems~\cite{Duivenvoorden_single-mode_2017}.

% \textcolor{red}{The reduced contrast is probably due to remaining interference patterns \textcolor{red}{(show Wigner functions right before measurement for this?)}.}

% \textit{-- Summary and Outlook}
\section{Summary and Outlook}
In conclusion, we presented an analytic theory for the geometric optimization of generalized echo protocols at any particle number and dephasing. The same optimization was already treated in~\cite{Gessner_metrological_2019} by Gessner, Smerzi and Pezz\`{e}. Using these results, we are able to give a comprehensive overview and characterization of the variational class of echo protocols in terms of the (un-)squeezing strengths. We find that only one new protocol exists. This protocol stands out as it exhibits Heisenberg scaling in the sensitivity even for strong dephasing during the OAT interaction. Remarkably, the effectively nonlinear readout performs almost as well as the quantum Fisher information in this case. 
Our results provide a route towards quantum enhanced measurements away from the typical squeezing regime or GHZ states with the measurement of spin projections only.
Beyond the scope of this paper, non-ideal signal and measurement directions or the impact of other imperfections such as noise during the phase imprint can be studied. 
The former is not expected to reduce sensitivity much for directions close to the optimal ones due to the predominant contribution of the largest singular value. 
The exact effect of the latter often depends on a number of other factors and requires additional modelling. For example a more precise modelling of the physical system at hand, the duration of the interrogation and the type and strength of the noise.
It is known that, beyond some critical ensemble size, noise during the signal acquisition reduces the scaling of quantum metrological amplification to the standard (classical) scaling of independent particles~\cite{Escher_general_2011, DemkowiczDobrzaski_elusive_2012}. This behavior is also expected for our protocols, most notably for the highly entangled states beyond the squeezing regime.
In most cases, however, it is the absolute performance at a given $N$ which matters and not the scaling.
We therefore believe that the tradeoff between quantum projection noise and technical noise, specific to each sensor, must be studied to understand in which cases entanglement is actually useful for metrological improvements~\cite{Schulte_spin_2019}.

\begin{acknowledgments}
We would like to thank C. Klempt for helpful discussions and J.P. Home for commenting on the possible connection to GKP protocols.
This work was funded by the Deutsche Forschungsgemeinschaft (DFG, German Research Foundation) under Germany’s Excellence Strategy – EXC-2123 QuantumFrontiers – 390837967 and CRC 1227 `DQ-mat' project A06.
The publication of this article was funded by the Open Access Fund of Leibniz Universit\"{a}t Hannover.
\end{acknowledgments}

\bibliography{references}

\onecolumn\newpage
\appendix

\section{Signal and noise from spin characteristic functions}\label{app:CharF}
The matrices $M$ and $Q$ can be conveniently evaluated from collective spin expectation values when transforming to a spherical basis ${S_+, S_z, S_-}$ and using the characteristic function approach of~\cite{Arecchi_atomic_1972}.
For the signal strength, we find
\begin{equation}
    M = \begin{pmatrix}
    \frac{1}{2} [n_1 + n_2] & 0 & 0 \\
    0 & \frac{1}{2} [n_1 - n_2] & n_3 \\
    0 & n_4 & 0
    \end{pmatrix}
\end{equation}
with
\begin{align*}
    n_1 &= \frac{N(N - 1)}{2} \sin\left(\frac{\mu-\nu}{2}\right) \cos^{N - 2}\left(\frac{\mu - \nu}{2}\right) \\
    n_2 &= -\frac{N(N-1)}{2} \sin\left(\frac{\mu - \nu}{2}\right) \cos^{N - 2}\left(\frac{\mu + \nu}{2}\right) \\
    n_3 &= -\frac{N}{2} \cos^{N - 1}\left(\frac{\mu}{2}\right) \\
    n_4 &= \frac{N}{2} \cos^{N - 1}\left(\frac{\nu}{2}\right) ~.
\end{align*}

For the spin covariances, the respective matrix is 
\begin{equation}
    Q = \begin{pmatrix}
    \frac{1}{2} [q_1 + q_2] - q_0^2 & 0 & 0 \\
    0 & \frac{1}{2} [q_1 - q_2] & q_3 \\
    0 & q_3 & q_4
    \end{pmatrix}
\end{equation}
with 
\begin{align*}
    q_0 &= \frac{N}{2}\, \cos^{N -1}\left(\frac{\nu}{2}\right), ~~ q_1 = \frac{N (N+1)}{4} \\
    q_2 &= \frac{N (N-1)}{4} \cos^{N - 2}\left(\nu\right) \\
    q_3 &= \frac{N (N-1)}{4} \sin\left(\frac{\nu}{2}\right) \cos^{N -2}\left(\frac{\nu}{2}\right) \\
    q_4 &= \frac{N}{4} ~.
\end{align*}
We show details of the derivation in the rest of this section and extend the calculations to collective and individual dephasing in the next section.

The spherical basis ${S_+, S_z, S_-}$ for a collective spin of length $S = N/2$ includes the angular momentum ladder operators $S_{\pm} = S_x \pm i\, S_y$.
Correspondingly, $S_x = \dfrac{1}{2} (S_+ + S_- )$, $S_y = \dfrac{1}{2i} (S_+ - S_- )$ and
\begin{equation}\label{eq:commutators}
    [S_+, S_-] = 2 S_z, \quad [S_z, S_{\pm}] = \pm S_{\pm}.
\end{equation}
In the following we aim at calculating expectation values with respect to spin coherent states $\vert \theta, \varphi \rangle = e^{-i\theta (S_x \sin \varphi - S_y \cos \varphi)} \vert -S \rangle_z$, where the special case $\varphi = 0$, $\theta = \pi/2$ is the initial state $\CSS$ of a standard Ramsey protocol after the first $\pi/2$-pulse.
Anti-normally ordered expectation values in the spherical basis, meaning that $S_-$ is always left of $S_z$ and both to the left of $S_+$, of the general form $\langle \theta, \varphi \vert S_-^c e^{\mathfrak{c} S_-} S_z^b e^{\mathfrak{b} S_z} S_+^a e^{\mathfrak{a} S_+} \vert \theta, \varphi \rangle$, with integers $a, b, c$ as well as arbitrary coefficients $\mathfrak{a}, \mathfrak{b}, \mathfrak{c}$, can be calculated via derivatives of a spin characteristic function~\cite{Arecchi_atomic_1972}.
Explicitly, 
\begin{equation}\label{eq:EVcharfunc}
    \langle \theta, \varphi \vert S_-^c e^{\mathfrak{c} S_-} S_z^b e^{\mathfrak{b} S_z} S_+^a e^{\mathfrak{a} S_+} \vert \theta, \varphi \rangle = \partial^a_{\alpha}\, \partial^b_{\beta}\, \partial^c_{\gamma}\,\, \charf (\alpha, \beta, \gamma) \big\rvert_{\alpha = \mathfrak{a}, \beta = \mathfrak{b}, \gamma = \mathfrak{c}}
\end{equation}
with the anti-normally ordered spin characteristic function~\cite{Arecchi_atomic_1972}
\begin{align}
    \charf (\alpha, \beta, \gamma) &= \langle \theta, \varphi \vert e^{\gamma S_-} e^{\beta S_z} e^{\alpha S_+} \vert \theta, \varphi \rangle \\
    &= \Big[e^{-\beta/2} \cos^2(\theta/2) + e^{\beta/2} \big(\sin(\theta/2) e^{-i\varphi} + \alpha \cos(\theta/2) \big) \big(\sin(\theta/2) e^{i\varphi} + \gamma \cos(\theta/2) \big) \Big]^{2 S}. \nonumber
\end{align}
Using the transformation matrix
\begin{equation}
    A = \begin{pmatrix}
     \frac{1}{2} & 0 & \frac{1}{2}\\
     \frac{1}{2i}& 0 & -\frac{1}{2i}\\
     0 & 1 & 0
    \end{pmatrix} \qquad \mathrm{and} \qquad     M_s = i \begin{pmatrix}
     \langle [S_+(\mu), S_+(\nu) ]\rangle & \langle [S_+(\mu), S_z(\nu) ]\rangle & \langle [S_+(\mu), S_-(\nu) ]\rangle\\
     \langle [S_z(\mu), S_+(\nu) ]\rangle & \langle [S_z(\mu), S_z(\nu) ]\rangle & \langle [S_z(\mu), S_-(\nu) ]\rangle\\
     \langle [S_-(\mu), S_+(\nu) ]\rangle & \langle [S_-(\mu), S_z(\nu) ]\rangle & \langle [S_-(\mu), S_-(\nu) ]\rangle
    \end{pmatrix},
\end{equation}
the matrix $M$ is related to its counterpart $M_s$ in the spherical basis via $M = A M_s A^T$.
In the same way the covariance variance $Q$ can be expressed as $Q = A Q_s A^T $ with
\begin{equation}
    Q_s = \frac{1}{2} \begin{pmatrix}
    \langle [S_+(\nu), S_+(\nu)]_+ \rangle & \langle [S_+(\nu), S_z(\nu)]_+ \rangle & \langle [S_+(\nu), S_-(\nu)]_+ \rangle\\
    \langle [S_z(\nu), S_+(\nu)]_+ \rangle & \langle [S_z(\nu), S_z(\nu)]_+ \rangle & \langle [S_z(\nu), S_-(\nu)]_+ \rangle\\
    \langle [S_-(\nu), S_+(\nu)]_+ \rangle & \langle [S_-(\nu), S_z(\nu)]_+ \rangle & \langle [S_-(\nu), S_-(\nu)]_+ \rangle
    \end{pmatrix} \nonumber -\vec{j}\, \vec{j}^T
\end{equation}
where $\vec{j} = \begin{pmatrix}
    \langle S_+(\nu) \rangle, 
    \langle S_z(\nu) \rangle,
    \langle S_-(\nu) \rangle
    \end{pmatrix}^T$ and $[\cdot,\cdot]_+$ denotes the anti-commutator.

To calculate the expectation values therein, they have to be brought into anti-normal order before using Eq.~\eqref{eq:EVcharfunc}.
For this we use that the transformed operators $S_{(+,z,-)}(\mu) = T_{\mu}^{\dagger} S_{(+,z,-)} T_{\mu}$ are $S_z(\mu)=S_z$ and 
\begin{equation}
    S_{\pm}(\mu) = e^{i\mu/2} S_{\pm} e^{\pm i \mu S_z} = e^{-i\mu/2} e^{\pm i \mu S_z} S_{\pm}.
\end{equation}
Furthermore, the transformations
\begin{align}
    &e^{i\mu S_z} S_{\pm} e^{-i \mu S_z} = e^{\pm i \mu} S_{\pm}, \qquad e^{-i\mu S_z} S_{\pm} e^{i \mu S_z} = e^{\mp i \mu} S_{\pm} \\
    &\Rightarrow S_+ e^{\pm i \mu S_z} = e^{\mp i \mu}\,\, e^{\pm i \mu S_z} S_+ , \qquad e^{\pm i \mu S_z} S_- = e^{\mp i \mu}\,\, S_- e^{\pm i \mu S_z}
\end{align}
are applied to obtain anti-normal ordering.
With this, we find for the first order moments
\begin{align}
    \langle S_+(\nu) \rangle &= e^{-i\nu/2} \langle e^{i \nu S_z} S_+ \rangle = e^{-i\nu/2} \partial_{\alpha} \charf \evat{\alpha = \gamma = 0, \beta = i\nu} 
    \\
    \langle S_z(\nu) \rangle &= \langle S_z \rangle = \partial_{\beta} \charf \evat{\alpha = \beta = \gamma = 0} \\
    \langle S_-(\nu) \rangle &= e^{i\nu/2} \langle S_- e^{-i \nu S_z} \rangle = e^{i\nu/2} \partial_{\gamma} \charf \evat{\alpha = \gamma = 0, \beta = -i\nu}
\end{align}
and all expectation values are with respect to the coherent spin state $\vert \theta, \varphi \rangle$.
For the symmetric second order moments:
\begin{align}
    \langle [S_+(\nu), S_+(\nu)]_+ \rangle &= 2 \langle S_+(\nu) S_+(\nu) \rangle \nonumber \\
    &= 2 e^{-i 2\nu} \langle e^{i 2\nu S_z} S_+^2 \rangle \nonumber \\
    &= 2 e^{-i 2\nu} \partial_{\alpha}\partial_{\alpha} \charf\evat{\alpha=\gamma=0, \beta=i 2\nu}
    \\
    \langle [S_+(\nu), S_z(\nu)]_+ \rangle &= \langle S_+(\nu) S_z + S_z S_+(\nu) \rangle \nonumber \\
    &= e^{-i\nu/2} \langle 2\, S_z e^{i\nu S_z} S_+ - e^{i\nu S_z} S_+ \rangle \nonumber \\
    &= e^{-i\nu/2} \Big\lbrace 2 \partial_{\beta} \partial_{\alpha} \charf\evat{\alpha=\gamma=0, \beta=i\nu} - \partial_{\alpha} \charf\evat{\alpha=\gamma=0, \beta=i\nu} \Big\rbrace
    \\
    \langle [S_+(\nu), S_-(\nu)]_+ \rangle &= \langle [S_+, S_-]_+ (\nu) \rangle \nonumber \\ 
    &= 2 \langle S_- S_+ + S_z \rangle \nonumber \\
    &= 2 \Big\lbrace \partial_{\gamma} \partial_{\alpha} \charf\evat{\alpha=\beta=\gamma=0} + \partial_{\beta} \charf\evat{\alpha=\beta=\gamma=0} \Big\rbrace 
\end{align}
\begin{align}
    \langle [S_z(\nu), S_+(\nu)]_+ \rangle &= \langle [S_+(\nu), S_z(\nu)]_+ \rangle
    \\
    \langle [S_z(\nu), S_z(\nu)]_+ \rangle &= 2 \langle S_z^2 \rangle \nonumber \\
    &= 2 \partial_{\beta} \partial_{\beta} \charf\evat{\alpha=\beta=\gamma=0}
    \\
    \langle [S_z(\nu), S_-(\nu)]_+ \rangle &= \langle S_z S_-(\nu) + S_-(\nu) S_z \rangle \nonumber \\ 
    &= e^{i\nu/2} \langle 2 S_- S_z e^{-i\nu S_z} - S_- e^{-i\nu S_z} \rangle \nonumber \\
    &= e^{i\nu/2} \Big\lbrace 2 \partial_{\beta} \partial_{\gamma} \charf\evat{\alpha=\gamma=0, \beta=-i\nu} - \partial_{\gamma} \charf\evat{\alpha=\gamma=0, \beta=-i\nu} \Big\rbrace 
\end{align}
\begin{align}
    \langle [S_-(\nu), S_+(\nu)]_+ \rangle &= \langle [S_+(\nu), S_-(\nu)]_+ \rangle
    \\
    \langle [S_-(\nu), S_z(\nu)]_+ \rangle &= \langle [S_z(\nu), S_-(\nu)]_+ \rangle
    \\
    \langle [S_-(\nu), S_-(\nu)]_+ \rangle &= 2 \langle S_-(\nu) S_-(\nu) \rangle \nonumber \\ 
    &= 2 e^{i 2\nu} \langle S_-^2 e^{-i2\nu S_z} \rangle \nonumber \\
    &= 2 e^{i 2\nu} \partial_{\gamma} \partial_{\gamma} \charf\evat{\alpha=\gamma=0, \beta=-i2\nu}
\end{align}
Finally, the moments for the commutators are:
\begin{align}
    \langle [S_+(\mu), S_+(\nu)] \rangle &= \langle S_+(\mu) S_+(\nu) - S_+(\nu) S_+(\mu) \rangle \nonumber \\
    &= e^{-i (\mu + \nu)/2} (e^{-i\nu} - e^{-i\mu}) \langle e^{i (\mu + \nu) S_z} S_+^2 \rangle \nonumber \\
    &= e^{-i (\mu + \nu)/2} (e^{-i\nu} - e^{-i\mu}) \partial_{\alpha}\partial_{\alpha} \charf\evat{\alpha=\gamma=0, \beta=i (\mu + \nu)}
    \\
    \langle [S_+(\mu), S_z(\nu)] \rangle &= \langle S_+(\mu) S_z - S_z S_+(\mu) \rangle \nonumber \\
    &= -e^{-i\mu/2} \langle e^{i\mu S_z} S_+ \rangle \nonumber \\
    &= -e^{-i\mu/2} \partial_{\alpha} \charf\evat{\alpha=\gamma=0, \beta=i\mu}
\end{align}
\begin{align}
    \langle [S_+(\mu), S_-(\nu)] \rangle &= \langle S_+(\mu) S_-(\nu) - S_-(\nu) S_+(\mu) \rangle \nonumber \\ 
    &= e^{-i(\mu-\nu)/2} \langle 2 S_z e^{i(\mu-\nu) S_z} + (e^{-i(\mu-\nu)} - 1)\, S_- e^{i(\mu - \nu) S_z} S_+ \rangle \nonumber \\
    &= e^{-i(\mu-\nu)/2} \Big\lbrace 2 \partial_{\beta} \charf\evat{\alpha=\gamma=0, \beta=i(\mu-\nu)} + (e^{-i(\mu-\nu)} - 1) \partial_{\gamma} \partial_{\alpha} \charf\evat{\alpha=\gamma=0, \beta=i(\mu-\nu)} \Big\rbrace \\
        \langle [S_z(\mu), S_+(\nu)] \rangle &= \langle S_z S_+(\nu) - S_+(\nu) S_z \rangle \nonumber \\
    &= - \langle [S_+(\nu), S_z] \rangle \nonumber \\
    &= e^{-i\nu /2} \partial_{\alpha} \charf\evat{\alpha=\gamma=0, \beta=i\nu}
\end{align}
\begin{align}
    \langle [S_z(\mu), S_z(\nu)] \rangle &= \langle [S_z, S_z] \rangle = 0
\end{align}
\begin{align}
    \langle [S_z(\mu), S_-(\nu)] \rangle &= \langle S_z S_-(\nu) - S_-(\nu) S_z \rangle \nonumber \\ 
    &= -e^{i\nu/2} \langle S_- e^{-i\nu S_z} \rangle \nonumber \\
    &= -e^{i\nu/2} \partial_{\gamma} \charf\evat{\alpha=\gamma=0, \beta=-i\nu}
\end{align}
\begin{align}
    \langle [S_-(\mu), S_+(\nu)] \rangle &= \langle S_-(\mu) S_+(\nu) - S_+(\nu) S_-(\mu) \rangle \nonumber \\
    &= e^{i(\mu-\nu)/2} \langle (1 - e^{i(\mu-\nu)})\, S_- e^{-i(\mu-\nu) S_z} S_+ - 2 S_z e^{-i(\mu - \nu) S_z} \rangle \nonumber \\
    &= e^{i(\mu-\nu)/2} \Big\lbrace (1 - e^{i(\mu-\nu)}) \partial_{\alpha} \partial_{\gamma} \charf\evat{\alpha=\gamma=0, \beta=-i(\mu-\nu)} - 2 \partial_{\beta} \charf\evat{\alpha=\gamma=0, \beta=-i(\mu-\nu)} \Big\rbrace
\end{align}
\begin{align}
    \langle [S_-(\mu), S_z(\nu)] \rangle &= - \langle [S_z, S_-(\mu)] \rangle \nonumber \\
    &= e^{i\mu /2} \langle S_- e^{-i \mu S_z} \rangle \nonumber \\
    &= e^{i \mu/2} \partial_{\gamma} \charf\evat{\alpha=\gamma=0, \beta=-i\mu}
\end{align}
\begin{align}
    \langle [S_-(\mu), S_-(\nu)] \rangle &= \langle S_-(\mu) S_-(\nu) - S_-(\nu) S_-(\mu) \rangle \nonumber \\ 
    &= e^{i (\mu + \nu)/2} (e^{i\mu} - e^{i\nu}) \langle S_-^2 e^{-i(\mu+\nu) S_z} \rangle \nonumber \\
    &= e^{i (\mu + \nu)/2} (e^{i\mu} - e^{i\nu}) \partial_{\gamma} \partial_{\gamma} \charf\evat{\alpha=\gamma=0, \beta=-i(\mu+\nu)}
\end{align}
With the characteristic function 
\begin{equation}
    \charf (0, \pi/2) = \Big[\frac{1}{2} e^{-\beta/2} + \frac{1}{2} e^{\beta/2} \big(1 + \alpha \big) \big(1 + \gamma \big) \Big]^{2 S}
\end{equation}
we then find
\begin{equation}
\begin{pmatrix}
    \langle S_+(\nu) \rangle \\
    \langle S_z(\nu) \rangle \\
    \langle S_-(\nu) \rangle
    \end{pmatrix} = 
\begin{pmatrix}
q_0 \\
0 \\
q_0
\end{pmatrix}
\end{equation}
with $q_0 = S\, \cos^{2 S -1}(\frac{\nu}{2}) $.
Likewise one finds
\begin{equation}
    Q_s = \begin{pmatrix}
    q_2 & i\, q_3 & q_1 \\
    i\, q_3 & q_4 &  -i\, q_3 \\
    q_1 & -i\, q_3 & q_2 
    \end{pmatrix} - \begin{pmatrix}
q_0 \\
0 \\
q_0
\end{pmatrix} \begin{pmatrix}
q_0, 
0, 
q_0
\end{pmatrix}
\end{equation}
and thus 
\begin{equation}
    Q = A Q_s A^T = \begin{pmatrix}
    \frac{1}{2} [q_1 + q_2]-q_0^2 & 0 & 0 \\
    0 & \frac{1}{2} [q_1 - q_2] & q_3 \\
    0 & q_3 & q_4
    \end{pmatrix}
\end{equation}
as well as
\begin{equation}
    M_s = \begin{pmatrix}
    n_2 & i\, n_3 & n_1 \\
    i\, n_4 & 0 &  -i\, n_4 \\
    n_1 & -i\, n_3 & n_2 
    \end{pmatrix}
    \qquad \mathrm{and} \qquad 
    M = A M_s A^T = \begin{pmatrix}
    \frac{1}{2} [n_1 + n_2] & 0 & 0 \\
    0 & \frac{1}{2} [n_1 - n_2] & n_3 \\
    0 & n_4 & 0
    \end{pmatrix}
\end{equation}
in accord to the results stated initially.

\section{Expectation values with dephasing}\label{app:noise}
This section contains details on calculating spin expectation values with dephasing noise.
The OAT dynamics in the case of collective dephasing is given by the master equation
\begin{equation}\label{eq:masterEqCollective}
    \dot{\rho} = -i [H, \rho] + \dephColl \Lcol[\rho]
\end{equation}
with $H = \chi S_z^2$, $\Lcol [\rho] = S_z \rho S_z - \frac{1}{2} S_z^2 \rho - \frac{1}{2} \rho S_z^2$ and the dephasing rate $\dephColl$.
Likewise, individual dephasing is described by the master equation
\begin{equation}\label{eq:masterEqIndiv}
    \dot{\rho} = -i [H, \rho] + \dephIndiv \Lind [\rho]
\end{equation}
with $H = \chi S_z^2$ and $\Lind [\rho] = \sum_{k=1}^N \sigma_z^{k} \rho \sigma_z^{k} - \rho$ where $\dephIndiv>0$ is the individual dephasing rate, equal for all particles.
The formal solution for collective dephasing is
\begin{equation}
    \rho (t) = T_{\mu} e^{\sigma \frac{\vert \mu \vert}{2} \Lcol}[\rho_0] T_{\mu}^{\dagger} = e^{\sigma \frac{\vert \mu \vert}{2} \Lcol}[T_{\mu} \rho_0 T_{\mu}^{\dagger}]
\end{equation}
from an initial state $\rho_0$ and with $\sigma = \vert \dephColl \vert / \vert \chi \vert$.
For individual dephasing we find
\begin{equation}
    \rho (t) = T_{\mu} e^{\Sigma \frac{\vert \mu \vert}{2} \Lind}[\rho_0] T_{\mu}^{\dagger} = e^{\Sigma \frac{\vert \mu \vert}{2} \Lind}[T_{\mu} \rho_0 T_{\mu}^{\dagger}]
\end{equation}
where $\Sigma = \vert \dephIndiv \vert / \vert \chi \vert$.
Expectation values of any operator $A$ are then
\begin{equation}
    \langle A \rangle = \tr{A \rho(t)}
    = \tr{A e^{\sigma \frac{\vert \mu \vert}{2} \mathcal{L}_k}[T_{\mu} \rho_0 T_{\mu}^{\dagger}]}
    = \tr{T_{\mu}^{\dagger} e^{\sigma \frac{\vert \mu \vert}{2} \mathcal{L}^{\dagger}_k} [A] T_{\mu}\, \rho_0} .
\end{equation}
Here, $\mathcal{L}^{\dagger}$ is the adjoint Lindblad operator, defined via
\begin{equation}
    \tr{A\, \mathcal{L}[B]} = \tr{\mathcal{L}^{\dagger}[A]\, B}
\end{equation}
so that
\begin{equation}
    \mathcal{L}^{\dagger}[A] = L^{\dagger} A L - \frac{1}{2} L^{\dagger} L A - \frac{1}{2} A L^{\dagger} L
\end{equation}
given
\begin{equation}
    \mathcal{L}[A] = L A L^{\dagger} - \frac{1}{2} L^{\dagger} L A - \frac{1}{2} A L^{\dagger} L.
\end{equation}
For both, collective and individual dephasing, this simplifies to $\Lcol^{\dagger} = \Lcol$ and $\Lind^{\dagger} = \Lind$.
For the protocols of the main text this allows to evaluate the expectation values
\begin{equation}
    \braket{S_-^{k_-} S_z^{k_z} S_+^{k_+}}\evat{\phi=0} = \tr{T_{\nu}^{\dagger} e^{\sigma \frac{\vert \mu \vert}{2} \mathcal{L}}\left[ e^{\sigma \frac{\vert \nu-\mu \vert}{2} \mathcal{L}} \left[S_-^{k_-} S_z^{k_z} S_+^{k_+}\right] \right] T_{\nu}\, \rho_0} 
\end{equation}
required for the spin covariance matrix as well as the slope
\begin{align*}
    \frac{\partial \braket{S_{\vec{m}}}}{\partial \phi}\evat{\phi = 0} 
    &= \frac{\partial}{\partial \phi}\left( \sum_k \tr{m_k S_k e^{\sigma \frac{\vert \nu - \mu \vert}{2} \mathcal{L}}\left[T_{\nu-\mu} e^{-i\phi \sum_l n_l S_l} e^{\sigma \frac{\vert \mu \vert}{2} \mathcal{L}}\left[T_{\mu} \rho_0 T_{\mu}^{\dagger}\right] e^{i\phi \sum_l n_l S_l} T_{\nu-\mu}^{\dagger} \right]} \right)_{\evat{\phi=0}} \\
    &= i \sum_{l,k} n_l M_{lk} m_k
\end{align*}
with 
\begin{equation}
    M_{l,k} = \tr{T_{\mu}^{\dagger} e^{\sigma \frac{\vert \mu \vert}{2} \mathcal{L}}\left[ [S_l, \, T_{\nu-\mu}^{\dagger} e^{\sigma \frac{\vert \nu - \mu \vert}{2} \mathcal{L}}[S_k] T_{\nu-\mu}] \right] T_{\mu} \rho}.
\end{equation}
(The same applies for individual dephasing with $\sigma \leftrightarrow \Sigma$ and $\mathcal{L} \leftrightarrow \mathcal{L}'$.)
The expectation values presented here can now be reduced to their noiseless version by explicitly evaluating the transformed operators.
At this point however we have to separate collective and individual dephasing. For collective dephasing the following holds:
First, it is clear that $ e^{\sigma \frac{\vert \mu \vert}{2} \mathcal{L}}[S_z] = S_z $ and from the commutation relations~\eqref{eq:commutators} it follows that $ e^{\sigma \frac{\vert \mu \vert}{2} \mathcal{L}}[S_{\pm}] = e^{-\sigma \frac{\vert \mu \vert}{4}} S_{\pm}$.
Repeated application of the commutation relations then also gives
\begin{align}
    e^{\sigma \frac{\vert \mu \vert}{2} \mathcal{L}}[S_{\pm}^2] &= e^{-\sigma \vert \mu \vert} S_{\pm}^2 \\
    e^{\sigma \frac{\vert \mu \vert}{2} \mathcal{L}}[S_{\pm} S_z] &= e^{-\sigma \frac{\vert \mu \vert}{4}} S_{\pm} S_z \\
    e^{\sigma \frac{\vert \mu \vert}{2} \mathcal{L}}[S_z S_{\pm}] &= e^{-\sigma \frac{\vert \mu \vert}{4}} S_z S_{\pm} \\
    e^{\sigma \frac{\vert \mu \vert}{2} \mathcal{L}}[S_{\pm} S_{\mp}] &= S_{\pm} S_{\mp}
\end{align}
which allows to express all expectation values to the ones with $\sigma = 0$ and appropriate exponential damping factors.
With this, we find
\begin{equation}
\begin{pmatrix}
    \langle S_+(\nu) \rangle \\
    \langle S_z(\nu) \rangle \\
    \langle S_-(\nu) \rangle
    \end{pmatrix} = 
\begin{pmatrix}
\tilde{q}_0 \\
0 \\
\tilde{q}_0
\end{pmatrix}
\end{equation}
with, again, $\tilde{q}_0 = e^{-\sigma \frac{\vert \nu - \mu \vert + \vert \mu \vert}{4}} S\, \cos^{2 S -1}(\frac{\nu}{2})$.\\
Likewise one finds
\begin{equation}
    \tilde{Q}_s = \begin{pmatrix}
    \tilde{q}_2 & i\, \tilde{q}_3 & \tilde{q}_1 \\
    i\, \tilde{q}_3 & \tilde{q}_4 &  -i\, \tilde{q}_3 \\
    \tilde{q}_1 & -i\, \tilde{q}_3 & \tilde{q}_2 
    \end{pmatrix} - \begin{pmatrix}
\tilde{q}_0 \\
0 \\
\tilde{q}_0
\end{pmatrix} \begin{pmatrix}
\tilde{q}_0, 
0, 
\tilde{q}_0
\end{pmatrix}
\end{equation}
and thus 
\begin{equation}
    \tilde{Q} = A \tilde{Q}_s A^T = \begin{pmatrix}
    \frac{1}{2} [\tilde{q}_1 + \tilde{q}_2]-\tilde{q}_0^2 & 0 & 0 \\
    0 & \frac{1}{2} [\tilde{q}_1 - \tilde{q}_2] & \tilde{q}_3 \\
    0 & \tilde{q}_3 & \tilde{q}_4
    \end{pmatrix}
\end{equation}
with
\begin{align}
    \tilde{q}_1 &= q_1 = S^2 + \frac{S}{2},\quad \tilde{q}_2 = e^{-\sigma (\vert \nu - \mu \vert + \vert \mu \vert)} \frac{S}{2} (2 S -1) \cos(\nu)^{2 S - 2}, \nonumber \\
    \tilde{q}_3 &= e^{-\sigma \frac{\vert \nu - \mu \vert + \vert \mu \vert}{4}} \frac{S}{2} (2 S -1) \cos(\frac{\nu}{2})^{2 S -2} \sin(\frac{\nu}{2}) \quad \mathrm{and} \quad \tilde{q}_4 = q_4 = \frac{S}{2}.
\end{align}
Finally,
\begin{equation}
    \tilde{M}_s = \begin{pmatrix}
    \tilde{n}_2 & i\, \tilde{n}_3 & \tilde{n}_1 \\
    i\, \tilde{n}_4 & 0 &  -i\, \tilde{n}_4 \\
    \tilde{n}_1 & -i\, \tilde{n}_3 & \tilde{n}_2 
    \end{pmatrix}
    \qquad \mathrm{and} \qquad 
    \tilde{M} = A \tilde{M}_s A^T = \begin{pmatrix}
    \frac{1}{2} [\tilde{n}_1 + \tilde{n}_2] & 0 & 0 \\
    0 & \frac{1}{2} [\tilde{n}_1 - \tilde{n}_2] & \tilde{n}_3 \\
    0 & \tilde{n}_4 & 0
    \end{pmatrix}
\end{equation}
with
\begin{equation}
    \tilde{n}_1 = e^{-\sigma \frac{\vert \nu - \mu \vert}{4}} S (2 S - 1) \sin(\frac{\mu-\nu}{2}) \cos(\frac{\mu - \nu}{2})^{2 S - 2} ,\quad \tilde{n}_2 = -e^{-\sigma (\frac{\vert \nu - \mu \vert}{4} + \vert \mu \vert)} S (2S - 1) \sin(\frac{\mu - \nu}{2}) \cos(\frac{\mu + \nu}{2})^{2S - 2},
\end{equation}
\begin{equation}
    \tilde{n}_3 = -e^{-\sigma \frac{\vert \mu \vert}{4}} S \cos(\frac{\mu}{2})^{2S - 1} \quad \mathrm{and} \quad \tilde{n}_4 = e^{-\sigma \frac{\vert \nu - \mu \vert + \vert \mu \vert}{4}} S \cos(\frac{\nu}{2})^{2S - 1}.
\end{equation}

A similar study shows that for individual dephasing the operators transform as
\begin{align}
    e^{\Sigma \frac{\vert \mu \vert}{2} \mathcal{L}'}[S_z] &= S_z \\
    e^{\Sigma \frac{\vert \mu \vert}{2} \mathcal{L}'}[S_{\pm}] &= e^{-\Sigma \vert \mu \vert} S_{\pm} \\
    e^{\Sigma \frac{\vert \mu \vert}{2} \mathcal{L}'}[S_{\pm}^2] &= e^{-2 \Sigma \vert \mu \vert} S_{\pm}^2 \\
    e^{\Sigma \frac{\vert \mu \vert}{2} \mathcal{L}'}[S_{\pm} S_z] &= e^{-\Sigma \vert \mu \vert} S_{\pm} S_z \\
    e^{\Sigma \frac{\vert \mu \vert}{2} \mathcal{L}'}[S_z S_{\pm}] &= e^{-\Sigma \vert \mu \vert} S_z S_{\pm} \\
    e^{\Sigma \frac{\vert \mu \vert}{2} \mathcal{L}'}[S_{\pm} S_{\mp}] &= e^{-2 \Sigma \vert \mu \vert} \left[ S_{\pm} S_{\mp} + \left(\frac{N}{2} \pm S_z \right) \left(e^{2 \Sigma \vert \mu \vert} - 1\right) \right]
\end{align}
With this, we find
\begin{equation}
\begin{pmatrix}
    \langle S_+(\nu) \rangle \\
    \langle S_z(\nu) \rangle \\
    \langle S_-(\nu) \rangle
    \end{pmatrix} = 
\begin{pmatrix}
q'_0 \\
0 \\
q'_0
\end{pmatrix}
\end{equation}
with $q'_0 = e^{-\Sigma (\vert \nu - \mu \vert + \vert \mu \vert)} q_0$.\\
Further
\begin{equation}
    Q'_s = \begin{pmatrix}
    q'_2 & i\, q'_3 & q'_1 \\
    i\, q'_3 & q'_4 &  -i\, q'_3 \\
    q'_1 & -i\, q'_3 & q'_2 
    \end{pmatrix} - \begin{pmatrix}
q'_0 \\
0 \\
q'_0
\end{pmatrix} \begin{pmatrix}
q'_0, 
0, 
q'_0
\end{pmatrix}
\end{equation}
and thus
\begin{equation}
    Q' = A Q'_s A^T = \begin{pmatrix}
    \frac{1}{2} [q'_1 + q'_2]-q_0^{\prime\, 2} & 0 & 0 \\
    0 & \frac{1}{2} [q'_1 - q'_2] & q'_3 \\
    0 & q'_3 & q'_4
    \end{pmatrix}
\end{equation}
with
\begin{align}
    q'_1 &= e^{-2 \Sigma (\vert \nu - \mu \vert + \vert \mu \vert)} q_1 + N/2 (1 - e^{-2 \Sigma (\vert \nu - \mu \vert + \vert \mu \vert)}),\quad q'_2 = e^{-2 \Sigma (\vert \nu - \mu \vert + \vert \mu \vert)} q_2, \nonumber \\
    q'_3 &= e^{- \Sigma (\vert \nu - \mu \vert + \vert \mu \vert)} q_3 \quad \mathrm{and} \quad q'_4 = q_4.
\end{align}
Finally,
\begin{equation}
    M'_s = \begin{pmatrix}
    n'_2 & i\, n'_3 & n'_1 \\
    i\, n'_4 & 0 &  -i\, n'_4 \\
    n'_1 & -i\, n'_3 & n'_2 
    \end{pmatrix}
    \qquad \mathrm{and} \qquad 
    M' = A M'_s A^T = \begin{pmatrix}
    \frac{1}{2} [n'_1 + n'_2] & 0 & 0 \\
    0 & \frac{1}{2} [n'_1 - n'_2] & n'_3 \\
    0 & n'_4 & 0
    \end{pmatrix}
\end{equation}
with
\begin{equation}
    n'_1 = e^{- \Sigma (\vert \nu - \mu \vert + 2 \vert \mu \vert)} ~ n_1 ,\quad n'_2 = e^{- \Sigma (\vert \nu - \mu \vert + 2 \vert \mu \vert)} ~ n_2,
\end{equation}
\begin{equation}
    n'_3 = e^{-\Sigma \vert \mu \vert} ~ n_3 \quad \mathrm{and} \quad n'_4 = e^{- \Sigma (\vert \nu - \mu \vert + \vert \mu \vert)} ~ n_4.
\end{equation}

\section{Review of quantum metrological sensitivity bounds with OAT}\label{app:QFI}
A fundamental bound to the achievable phase sensitivity is given by the quantum Cram\'{e}r-Rao bound~\cite{Helstrom_detection_1969}
\begin{equation}
    \Delta \phi \geq \frac{1}{\sqrt{\mathcal{F}_Q [\rho_0, H]}}
\end{equation}
this implies
\begin{equation}
    \left(\Delta \phi^{-1}\right)^2 \leq \mathcal{F}_Q [\rho_0, H]
\end{equation}
where $\mathcal{F}_Q [\rho_0, H]$ denotes the quantum Fisher information of a state $\rho_0$ onto which the phase is imprinted by the Hermitian operator $H$.
In this case, using the spectral decomposition $\rho_0 = \sum_{\kappa} p_{\kappa} \ket{\kappa} \bra{\kappa}$ with eigenvalues $p_{\kappa} \geq 0$ and associated eigenvectors $\bra{\kappa}$, the quantum Fisher information can be expressed as~\cite{Braunstein_statistical_1994, Pezze_quantumMetrology_2018}
\begin{equation}\label{eq:QFIhermitian}
    \mathcal{F}_Q [\rho_0, H] = 2 \sum_{\substack{\kappa, \kappa' \\ p_{\kappa} + p_{\kappa'} >0}} \frac{(p_{\kappa} - p_{\kappa'})^2}{p_{\kappa} + p_{\kappa'}} \vert \bra{\kappa'} H \ket{\kappa} \vert^2 .
\end{equation}
Here we only consider the case of initial states created via OAT, with or without collective dephasing, and unitary rotations with $H = S_{\vec{n}}$ imprinting the phase.
Without collective dephasing noise, i.e. $\rho_0 = T_{\mu} \CSS \braCSS T_{\mu}^{\dagger} $, it is known that the largest quantum Fisher information is~\cite{Pezze_quantumMetrology_2018}
\begin{equation}
    \max_{\vec{n}} \mathcal{F}_Q [\rho_0, S_{\vec{n}}] = \max\left\lbrace N+ \frac{N(N-1)}{4}\left(A + \sqrt{A^2 + B^2}\right),\,\, N^2 \left(1- \cos^{2N-2}(\mu/2) \right) - \frac{N (N-1) A}{2} \right\rbrace
\end{equation}
with $A = 1 - \cos^{N-2}(\mu),\,\, B = 4 \sin(\mu/2) \cos^{N-2}(\mu/2)$.
When including dephasing, in the form of the master equation~\eqref{eq:masterEq}, the density matrix can be expressed in the Dicke basis as
\begin{equation}\label{eq:rhoInDickeBasis}
    \rho = \sum_{m, m' = -N/2}^{N/2} e^{-i (m^2 - m'^2) \mu/2 - \sigma (m-m')^2 \vert \mu \vert/4}\, c_m (\theta, \varphi) c_{m'}^* (\theta, \varphi) \, \ket{m} \bra{m'}
\end{equation}
with the coefficients
\begin{equation}
    c_m (\theta, \varphi) = \binom{N}{N/2 + m}^{1/2} \sin^{N/2 + m}(\theta/2) \cos^{N/2 - m}(\theta/2) e^{-i (N/2 + m)\varphi}
\end{equation}
of the coherent spin state $\ket{\varphi, \theta}$ in the Dicke basis~\cite{Arecchi_atomic_1972}.
The quantum Fisher information for the dephased initial state can then be evaluated by numerically diagonalizing $\rho$ based on Eq.~\ref{eq:rhoInDickeBasis} and optimizing the right hand side of Eq.~\ref{eq:QFIhermitian}, with $H = S_{\vec{n}}$, over all directions $\vec{n}$.

%\begin{figure*}[tbp]
%	\centering
%	\includegraphics[width=\linewidth]{all_qfi_vgl}
%	\caption{Extended view of the comparison to QFI, for $N= 2-1024$.}
%	\label{fig:all_qfi}
%\end{figure*}

%\section{Details on Wigner functions}\label{app:Wigner}
%\begin{itemize}
%    \item Wigner function in terms of spherical harmonics (Ref. Agarwal, Schleich)
%    \item Coefficients for the state (with small rotation approximation)
%    \item Coefficients for the measurement operator
%\end{itemize}

\pagebreak

\begin{figure*}[tbp]
	\centering
	\includegraphics[width=\linewidth]{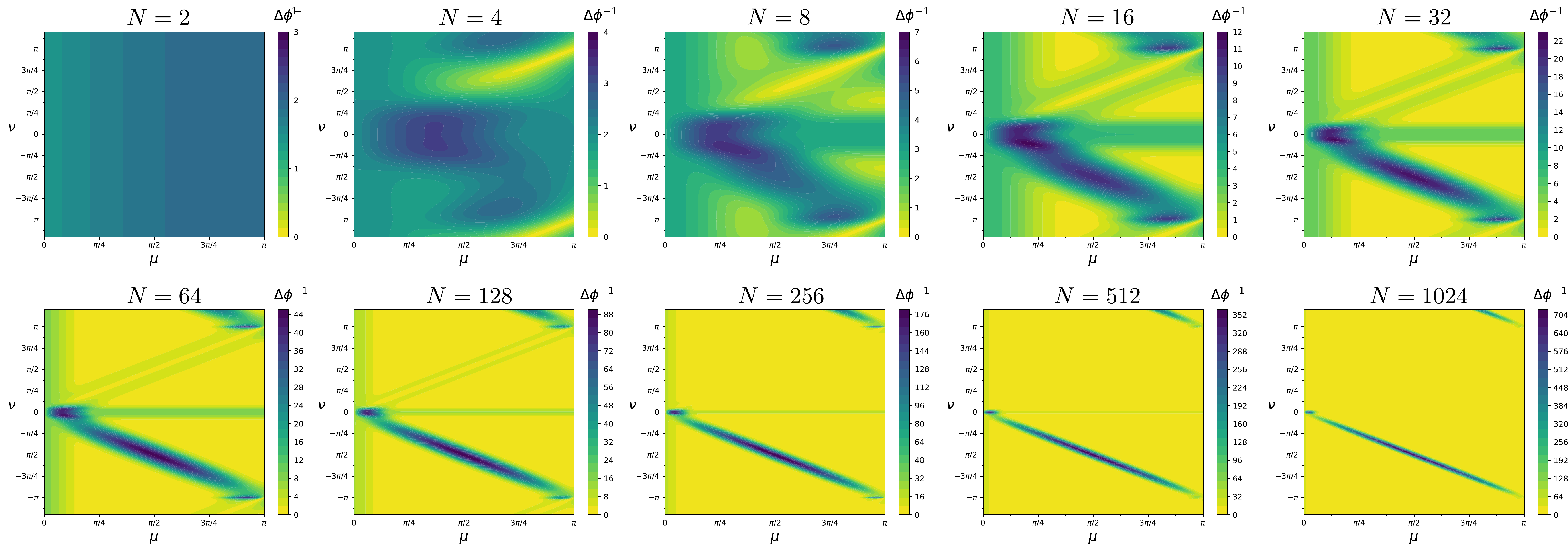}
	\caption{Extended view of the enhancement landscapes, for $N= 2-1024$.
	}
	\label{fig:all_landscapes}
\end{figure*}

\begin{figure*}[tbp]
	\centering
	\includegraphics[width=\linewidth]{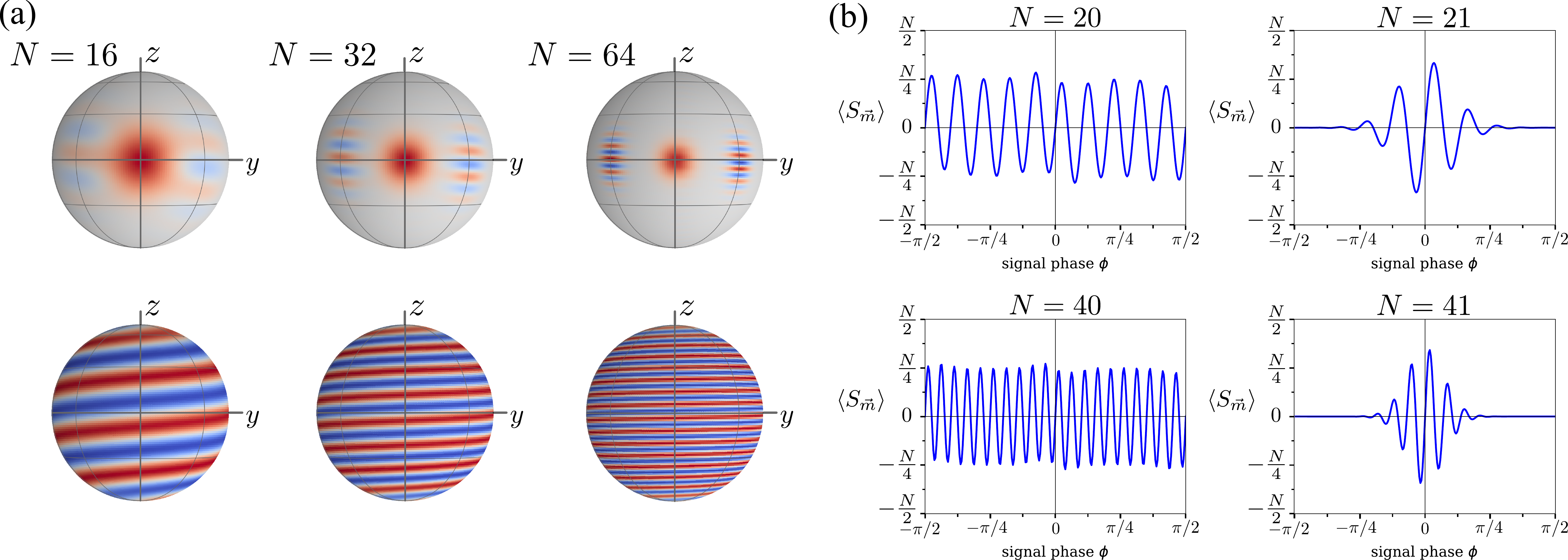}
	\caption{(a) Finer features appearing on the Wigner functions for the state $\ket{\psi} T_{-\mu} R_y(\phi) T_{\mu} \CSS $ (upper row) and the measurement operator $ T_{\nu}^{\dagger} S_y T_{\nu} $ (lower row). (b) OUT signals beyond $\phi \ll 1$. Even particle numbers show an $N$-fold increased oscillation whereas odd particle numbers experience a strong signal for small phases and subsequent attenuation.
	}
	\label{fig:wigner_functions}
\end{figure*}

\end{document}